\documentclass[twocolumn,letter]{jpsj3}

\usepackage{graphicx}
\usepackage{dcolumn}
\usepackage{bm}
\usepackage{color}
\usepackage{ulem}

\begin{document}

\title{Two-dimensional XY-type Magnetic Properties of \\ Locally Noncentrosymmetric Superconductor CeRh$_2$As$_2$}

\author{Shunsaku~Kitagawa$^{1,}$\thanks{E-mail address: kitagawa.shunsaku.8u@kyoto-u.ac.jp},
Mayu~Kibune$^{1}$, 
Katsuki~Kinjo$^{1}$, 
Masahiro~Manago$^{1,}$
\thanks{Present address:Department of Physics and
Materials Science, Graduate School of Natural Science and Technology,
Shimane University, Matsue 690-8504, Japan}, 
Takanori~Taniguchi$^{1,}$
\thanks{Present address:Institute for Materials Research, Tohoku University, Katahira, Sendai 980-8577, Japan}, 
Kenji~Ishida$^{1}$, 
Manuel~Brando$^{2}$, 
Elena~Hassinger$^{2}$, 
Christoph~Geibel$^{2}$, 
Seunghyun~Khim$^{2}$}
\inst{$^1$Department of Physics, Graduate School of Science, Kyoto University, Kyoto 606-8502, Japan \\
$^2$Max Planck Institute for Chemical Physics of Solids, D-01187 Dresden, Germany}

\date{\today}

\abst{
We performed $^{75}$As-NMR measurements to investigate the normal-state magnetic properties of CeRh$_2$As$_2$, a recently-discovered heavy-fermion superconductor.
The magnitude and temperature dependence of the Knight shift at the As(2) site indicate easy-plane-type magnetic anisotropy in CeRh$_2$As$_2$.
With regard to spin fluctuations, the temperature dependence of the nuclear spin-lattice relaxation rate $1/T_1$ arising from the 4$f$ electrons decreases from high-temperature constant behavior on cooling at $\sim$ 40~K, which is a typical behavior of heavy-fermion systems.  
In addition, $1/T_1$ becomes constant at low temperatures, suggesting spatially two-dimensional antiferromagnetic fluctuations.
Two-dimensional magnetic correlations in the real space are quite rare among heavy-fermion superconductors, and they may be a key factor in the unique superconducting multi phase in CeRh$_2$As$_2$.
}

\maketitle

Unconventional superconductivity is mostly considered to be mediated by magnetic fluctuations\cite{P.Monthoux_Nature_2007,C.Pfleiderer_RMP_2009,K.Ishida_JPSJ_2009}.
Thus, investigating magnetic properties in the normal state provides important information about the superconducting (SC) mechanism\cite{Y.Nakai_PRB_2013,K.Ishida_PRB_2021}.

CeRh$_{2}$As$_{2}$ is a recently discovered heavy-fermion superconductor whose $T_{\rm SC}$ is approximately 0.3~K\cite{S.Khim_Science_2021}.
The heavy-fermion superconductivity is characterized by a broad maximum in resistivity at $T_{\rm coh} \sim$ 40~K and a large specific-heat jump at $T_{\rm SC}$.
In many Ce-based heavy-fermion compounds, superconductivity appears in a narrow region near the antiferromagnetic (AFM) quantum critical point as a function of tuning parameters such as pressure, and it is considered to be mediated by AFM fluctuations\cite{C.Pfleiderer_RMP_2009}.
Even in CeRh$_2$As$_2$, AFM fluctuations are expected to play an important role.
Further, CeRh$_2$As$_2$ exhibits additional features that are not seen in other systems.

\begin{figure}[!tb]
\includegraphics[width=8.5cm,clip]{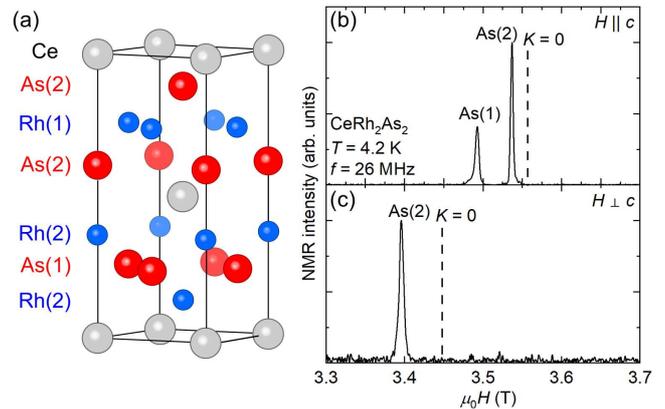}
\caption{(a)Crystal structure of CeRh$_{2}$As$_{2}$.
NMR spectrum arising from the central transition (-1/2 $\leftrightarrow$ 1/2) of CeRh$_2$As$_2$ at 4.2~K for (b) $H \parallel c$ and (c) $H \perp c$.
The dashed lines indicate the position of $K = 0$ when the contribution of the nuclear quadrupole interaction is taken into account.
}
\label{Fig.1}
\end{figure}

One such feature is the lack of local inversion symmetry in Ce layeres with global inversion symmetry.
The crystal structure is of the tetragonal CaBe$_{2}$Ge$_{2}$-type with space group $P4/nmm$ (No.129, $D_{4h}^7$)\cite{R.Madar_JLCM_1987}.
CeRh$_{2}$As$_{2}$ has two crystallographically inequivalent As and Rh sites; As(1) [Rh(1)] is tetrahedrally coordinated by Rh(2) [As(2)], as shown in Fig.~\ref{Fig.1} (a).
The crystal structure looks similar to that of the typical heavy-fermion superconductor CeCu$_{2}$Si$_{2}$ (i.e., a ThCr$_2$Si$_2$-type structure)\cite{F.Steglich_PRL_1979}.
Both compounds have inversion symmetry in crystal structures.
However, the stacking order of the block layers is different.
In CeRh$_2$As$_2$, a Ce layer is located between two different block layers (Rh-As-Rh and As-Rh-As), and there is no inversion center at the Ce site.
In contrast, in CeCu$_2$Si$_2$, a Ce layer is located between identical block layers (Si-Cu-Si).
Further, in Ce$M$Si$_{3}$ ($M$ = Rh or Ir), a Ce layer is sandwiched between Si-Si-$M$ block layers and global inversion symmetry is broken\cite{N.Kimura_PRL_2005}, making CeRh$_{2}$As$_{2}$ a good comparison to investigate how a lack of local inversion symmetry affects unconventional SC and magnetic states.
For superconductors with no inversion center at particular atomic sites with global inversion symmetry, theoretical studies have noted that odd-parity SC states such as the pair density wave state can be stabilized under magnetic fields even if a pairing interaction of the spin-singlet channel only exists\cite{T.Yoshida_PRB_2012}.
In fact, a first-order SC--SC phase transition at approximately 4~T was reported for $H \parallel c$\cite{S.Khim_Science_2021}; this can be interpreted as a phase transition from a low-field even-parity state to a high-field odd-parity state inside the SC phase\cite{E.G.Schertenleib_PRR_2021,K.Nogaki_PRR_2021,D.Mockli_PRR_2021}.

Another feature of CeRh$_2$As$_2$ is the nonmagnetic and magnetic phase transition just above and below $T_{\rm SC}$, respectively\cite{S.Khim_Science_2021,M.Kibune_PRL_2021}.
The specific heat shows a large anomaly at $T_{\rm SC}$ and a rather weak one at $T_0 \sim$ 0.4~K.
Magnetization and AC susceptibility measurements show no anomaly at $T_0$, and the $T_0$ anomaly increases with magnetic fields applied perpendicular to the $c$ axis, suggesting that it is not a simple magnetic order.
On the basis of recent renormalized band structure calculations, a quadrupole-density-wave transition at $T_0$ has been proposed\cite{D.Hafner_arXiv_2021}.
In addition to the nonmagnetic transition at $T_0$, our nuclear quadrupole resonance (NQR) measurements revealed an AFM order inside the SC phase\cite{M.Kibune_PRL_2021}.
The NQR spectrum of two crystallographically inequivalent As sites showed site-dependent linewidth broadening below $T_{\rm N}$ = 0.25~K.
Because the internal magnetic field can be canceled out at the symmetric site due to the superposition of the transferred fields from different Ce neighbors, the site-dependent linewidth broadening is the evidence of AFM transition.
Furthermore, the orbital order due to the existence of two nonequivalent positions of Ce atoms was theoretically predicted\cite{A.Ptok_PRB_2021}.
Therefore, CeRh$_{2}$As$_{2}$ is a promising system to study how the absence of local inversion symmetry induces or influences unconventional nonmagnetic, AFM and SC states, as well as their interaction.

In this study, we performed $^{75}$As-NMR measurements to investigate the normal-state magnetic properties of CeRh$_2$As$_2$.
The temperature dependence of the Knight shift and $1/T_1$ indicates spatially two-dimensional magnetic fluctuations with easy-plane ($XY$)-type anisotropy in CeRh$_2$As$_2$.
Two-dimensional magnetic correlations in the real space are quite different from those of other heavy-fermion superconductors, as a result of which it exhibits a unique SC state.

Single crystals of CeRh$_{2}$As$_{2}$ and a non-magnetic reference compound LaRh$_2$As$_2$ were grown by the Bi flux method\cite{S.Khim_Science_2021}.
The bulk magnetic susceptibility was measured using a commercial magnetometer with a superconducting
quantum interference device (Quantum Design, MPMS).
For NMR measurements, we used a split SC magnet, that generated a horizontal field and combined it with a single-axis rotator to apply a magnetic field parallel or perpendicular to the $c$ axis.
The $^{75}$As-NMR spectra (nuclear spin $I~=~3/2$, nuclear gyromagnetic ratio $\gamma/2\pi~=~7.29$~MHz/T, and natural abundance 100\%) were obtained as a function of the magnetic field at a fixed frequency ($\sim$ 26~MHz).
Figs.~\ref{Fig.1} (b) and \ref{Fig.1} (c) show typical NMR spectra for $H \parallel c$ and $H \perp c$, respectively.
To ascribe the NMR peaks to the As sites and estimate the Knight shift $K_i$ ($i = c$ and $\perp$), we computed the resonance magnetic field by diagonalizing the following Hamiltonian.
\begin{align}
\mathcal{H} = - &\gamma \hbar (1 + K_i)I\cdot H + \notag\\ &\frac{h\nu_{\mathrm{Q}}}{6}\left[3I_{z}^{2}-I\left(I+1\right)+\frac{\eta}{2}\left(I_{+}^{2}+I_{-}^{2}\right)\right],
\label{eq.1}
\end{align}
where $h$, $\nu_{\mathrm{Q}}=\frac{3heQV_{zz}}{2I\left(2I-1\right)}$, and $\eta = \left|\frac{V_{yy}-V_{xx}}{V_{zz}}\right|$ are the Planck constant, NQR frequency, and asymmetry parameter, respectively.
$\eta$ is zero at each As site because of the 4-fold symmetry of the atomic position.
Because $\nu_{\mathrm{Q}}$ of the As(1) site ($\sim$~30~MHz) is larger than that of the As(2) site ($\sim$~10~MHz)\cite{M.Kibune_PRL_2021}, the site assignments can be made from the angular dependence of the NMR peaks.
The temperature dependence of $\nu_{\mathrm{Q}}$ was determined from NQR measurements and the interval between NMR satellite signals.
For $H \parallel c$, because the contribution of the nuclear quadrupole interaction can be ignored at the center peak, the Knight shift is simply determined by $K = (H_0-H)/H$. 
Here, $H_0$ is the reference field.
It was estimated using $^{63}$Cu ($^{63}\gamma_n/2\pi = 11.285$~MHz/T) and $^{65}$Cu ($^{65}\gamma_n/2\pi = 12.08$~MHz/T) NMR signals from the NMR coil. 
The nuclear spin-lattice relaxation rate $1/T_1$ was determined by fitting the time variation of the nuclear magnetization probed with the spin-echo intensity after saturation to a theoretical function for $I$ = 3/2\cite{A.Narath_PR_1967,D.E.MacLaughlin_PRB_1971}.

\begin{figure}[!tb]
\includegraphics[width=8.2cm,clip]{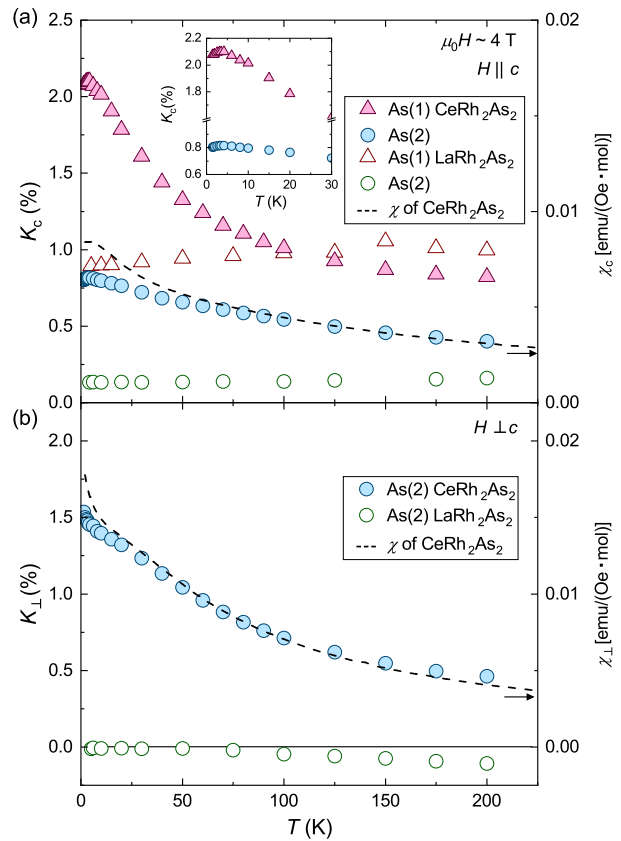}
\caption{Temperature dependence of Knight shift $K_i$ ($i = c$, and $\perp$)for (a) $H \parallel c$ and (b) $H \perp c$ in CeRh$_{2}$As$_{2}$.
For comparison, the temperature dependence of $K_i$ in LaRh$_{2}$As$_{2}$ and magnetic susceptibility $\chi_i$ are also depicted.
Measurements were performed at $\sim 4$~T.
The $K = 0$ line is indicated by a solid line.  
(Inset) Enlarged view in low-temperature range.
}
\label{Fig.2}
\end{figure}

\begin{table*}[!tb]
\caption{Estimated hyperfine coupling constants $A_{\rm hf}$ and $K_{\rm orb}$.
Temperatures in parentheses indicate the fitted temperature range.}
    \footnotesize
    \centering
    \begin{tabular}{cccccc}
    \hline
        \begin{tabular}{c}
            Site
        \end{tabular} & 
        \begin{tabular}{c}
            Direction of $H$
        \end{tabular} & 
        \begin{tabular}{c}
            $A_{\rm hf}$ (T/$\mu_{\rm B}$) \\ in low $T$  
        \end{tabular} & 
        \begin{tabular}{c}
            $A_{\rm hf}$ (T/$\mu_{\rm B}$)  \\ in high $T$
        \end{tabular} & 
        \begin{tabular}{c}
            $K_{\rm orb}$ (\%) \\ in low $T$  
        \end{tabular} & 
        \begin{tabular}{c}
            $K_{\rm orb}$ (\%)  \\ in high $T$
        \end{tabular} \\
    \hline \hline
    As(1) & $H \parallel c$ & 
        \begin{tabular}{c}
            1.55  \\ (2-100~K)
        \end{tabular} &
        \begin{tabular}{c}
            0.46  \\ (125-200~K)
        \end{tabular} &
        \begin{tabular}{c}
            -0.24  \\ (2-100~K)
        \end{tabular} &
        \begin{tabular}{c}
            0.57  \\ (125-200~K)
        \end{tabular} \\
    \hline
    As(2) & $H \parallel c$ &  
        \begin{tabular}{c}
            0.27  \\ (2-30~K)
        \end{tabular} &
        \begin{tabular}{c}
            0.55  \\ (40-200~K)
        \end{tabular} &
        \begin{tabular}{c}
            0.41  \\ (2-30~K)
        \end{tabular} &
        \begin{tabular}{c}
            0.10  \\ (40-200~K)
        \end{tabular} \\
    & $H \perp c$ &
        \begin{tabular}{c}
            0.16  \\ (2-10~K)
        \end{tabular} &
        \begin{tabular}{c}
            0.50  \\ (15-200~K)
        \end{tabular} &
        \begin{tabular}{c}
            0.98  \\ (2-10~K)
        \end{tabular} &
        \begin{tabular}{c}
            0.09  \\ (15-200~K)
        \end{tabular}\\
    \hline
    \end{tabular}
    \label{tab1}
\end{table*}

Figures~\ref{Fig.2} (a) and \ref{Fig.2} (b) show the temperature dependence of the Knight shift in CeRh$_{2}$As$_{2}$ and LaRh$_{2}$As$_{2}$.
$K_{c}$ and $K_{\perp}$ denote the Knight shift for $H \parallel c$ and $H \perp c$, respectively.
Because $\nu_{\mathrm{Q}}$ at the As(1) site is large ($\sim$ 30~MHz), $K_{\perp}$ at the As(1) site could not be measured in the present measurement setup.
For comparison, the temperature dependence of the magnetic susceptibility $\chi_i$ ($i = c$ and $\perp$) is also shown.
The Knight shift is proportional to the magnetic susceptibility as given by,
\begin{align}
K_{i} = A_{{\rm hf},i} \chi_{{\rm spin},i} + K_{{\rm orb},i},
\label{eq.K}
\end{align}
where $A_{{\rm hf},i}$, $\chi_{{\rm spin},i}$, and $K_{{\rm orb},i}$ are respectively the hyperfine coupling constant, spin susceptibility, and orbital part of the Knight shift in each direction (which is generally temperature independent).
$K_i$ in CeRh$_{2}$As$_{2}$ for both magnetic field directions increased on cooling.
While $K_c$ shows a broad maximum at 3~K [see the inset of Fig.~\ref{Fig.2} (a)], $K_{\perp}$ continued to increase down to 1.5~K.
This behavior is consistent with the bulk magnetic susceptibility, which shows the $XY$-type magnetic anisotropy.
Most heavy-fermion superconductors such as CeCu$_{2}$Si$_{2}$\cite{T.Ohama_JPSJ_1995,S.Kitagawa_PRB_2017} and CeIrSi$_{3}$\cite{Y.Okuda_JPSJ_2007,H.Mukuda_PRL_2010} have this kind of $XY$-type magnetic anisotropy.
Unlike in the case of CeRh$_{2}$As$_{2}$, $K_i$ in LaRh$_{2}$As$_{2}$ slightly decreased on cooling.
In conventional metals, $K_i$ remains constant with changes in temperature.
The weak temperature-dependent $K_i$ in LaRh$_{2}$As$_{2}$ can be explained by the increase in the density of states due to thermally excited quasi-particles.
This suggests the presence of a peak of the density of states near the Fermi energy $E_{\rm F}$, as in the case of iron-based superconductors\cite{F.L.Ning_JPSJ_2009,S.Kitagawa_PRB_2010}.
We note that $K_c$ at the As(1) site in LaRh$_{2}$As$_{2}$ is larger than that in CeRh$_{2}$As$_{2}$.
The reason for this is unclear, and this issue requires further study.

\begin{figure}[!b]
\includegraphics[width=8.2cm,clip]{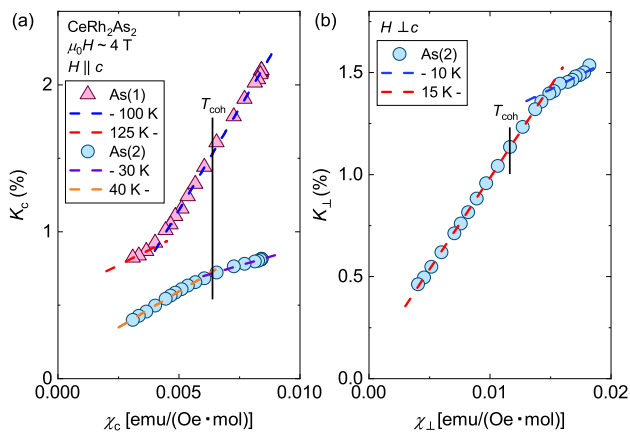}
\caption{$K$--$\chi$ plot for (a) $H \parallel c$ and (b) $H \perp c$ in CeRh$_{2}$As$_{2}$.
The broken lines indicate the linear fitting results.
The solid lines indicate $\chi_i$ value at $T_{\rm coh} \sim$  40~K.}
\label{Fig.3}
\end{figure}

\begin{figure*}[!tb]
\includegraphics[width=17cm,clip]{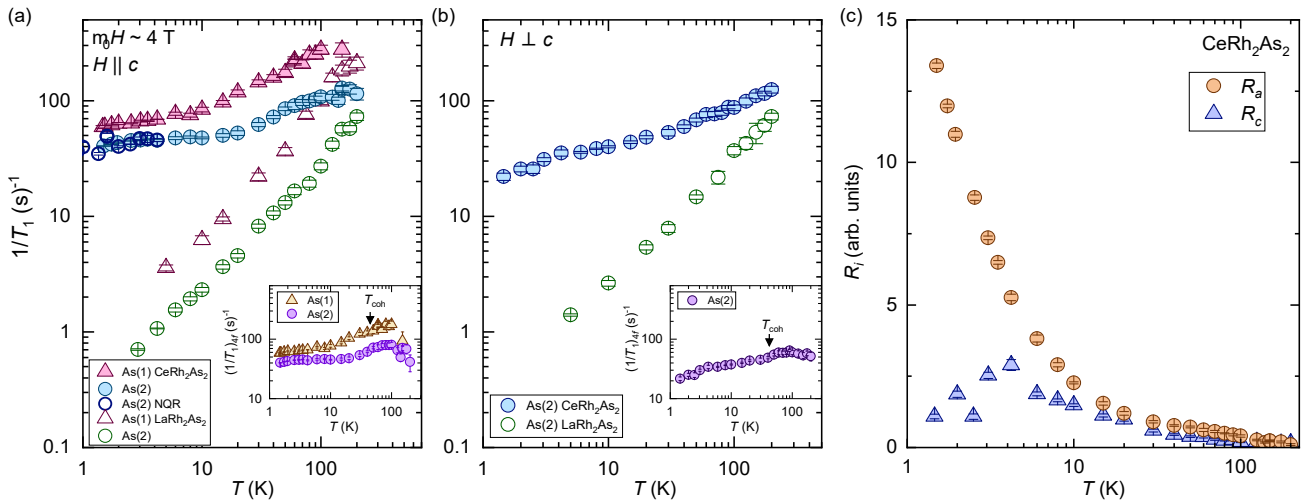}
\caption{Information about magnetic fluctuations.
Temperature dependence of $1/T_1$ for (a) $H \parallel c$ and (b) $H \perp c$ in CeRh$_{2}$As$_{2}$ and LaRh$_{2}$As$_{2}$.
The temperature dependence of $1/T_1$ in NQR measurements is also depicted.
The insets show the contribution from Ce 4$f$ moments $(1/T_1)_{4f}$. 
The arrows indicate $T_{\rm coh}$.
(c) Temperature dependence of $R_a$ and $R_c$ at As(2) site in CeRh$_2$As$_2$.
}
\label{Fig.4}
\end{figure*}

The hyperfine coupling constant $A_{{\rm hf},i}$ can be determined by the slope of the $K$--$\chi$ plot, as shown in Figs.~\ref{Fig.3} (a) and \ref{Fig.3} (b).
This slope changes between regions where $\chi_i$ is small and large.
This behavior is often observed in Ce-based heavy-fermion systems, and it is attributed to the development of the heavy-fermion state owing to $c$--$f$ hybridization\cite{T.Ohama_JPSJ_1995,N.J.Curro_PRB_2001}.
Indeed, the position of the slope change is roughly coincident with $T_{\rm coh}$, as indicated by the solid lines in Figs.~\ref{Fig.3} (a) and \ref{Fig.3} (b).
Table~\ref{tab1} summarizes the estimated hyperfine coupling constants and $K_{\rm orb}$.
$A_{{\rm hf},c}$ at the As(2) site becomes larger on cooling, whereas $A_{{\rm hf},c}$ and $A_{{\rm hf},\perp}$ at the As(1) site become smaller on cooling.
Assuming the classical dipole interaction, $A_{{\rm hf},i}$ was calculated to be 0.02-0.06 T/$\mu_{\rm B}$, which is of one order of magnitude smaller than the experimentally estimated $A_{{\rm hf},i}$.
Therefore, the transferred hyperfine interaction is dominant in CeRh$_{2}$As$_{2}$.

Next, we discuss the spin dynamics in CeRh$_2$As$_2$.
Figures~\ref{Fig.4} (a) and \ref{Fig.4} (b) show the temperature dependence of $1/T_1$.
$1/T_1$ of CeRh$_2$As$_2$ showed a complex temperature dependence; specifically, it decreased slowly at high temperatures, decreased significantly below $\sim$ 100~K, and finally became constant below 10~K.
In contrast, $1/T_1$ of LaRh$_2$As$_2$ is roughly proportional to temperature in both magnetic-field directions; this is the typical behavior of conventional metals.
To extract the contribution of Ce 4$f$ moments $(1/T_1)_{4f}$, $1/T_1$ of LaRh$_2$As$_2$ was subtracted from that of CeRh$_2$As$_2$ as shown in the inset of Fig.~\ref{Fig.4}.
Although the data above 100~K is scattered owing to an experimental error, the constant $(1/T_1)_{4f}$ at high temperatures reflects the localized 4$f$ moments.
On cooling, $(1/T_1)_{4f}$ decreased below 40~K, which is consistent with the broad maximum of the electrical resistivity at $T_{\rm coh}$ $\sim$ 40~K\cite{S.Khim_Science_2021}.
Thus, this decrease in $(1/T_1)_{4f}$ is related to the development of the heavy-fermion state.
On further cooling, $(1/T_1)_{4f}$ becomes constant again, unlike in the case of other heavy-fermion systems.
Because the Ce 4$f$ moments have an itinerant character at low temperatures, the temperature dependence of $1/T_1$ can be understood by the self-consistent renormalization (SCR) theory\cite{T.Moriya_JPSJ_1994,Y.Nakai_PRB_2013,J.L.Sarrao_PhysicaC_2015}.
According to the SCR theory, a constant $1/T_1$ implies spatially two-dimensional AFM fluctuations, as is often observed in high-$T_{\rm SC}$ cuprates\cite{S.Ohsugi_JPSJ_1991,K.Ishida_JPSJ_1993} and iron-based superconductors\cite{Y.Nakai_PRL_2010,S.Kitagawa_PRB_2019}, but not in heavy-fermion superconductors.
In CeCu$_2$Si$_2$ and CeIrSi$_3$, $1/T_1$ is proportional to $T$\cite{S.Kitagawa_PRB_2017} and $\sqrt{T}$\cite{H.Mukuda_PRL_2008} just above $T_{\rm SC}$, indicating a weakly correlated metallic state and the development of spatially three-dimensional AFM fluctuations, respectively.
In addition, the $1/T_1$ at low temperatures in the NQR measurement is almost the same as that in the NMR measurement, as shown in Fig.~\ref{Fig.4} (a), and thus, the two-dimensional AFM fluctuations are related to the AFM order inside the SC phase rather than the magnetic field-induced magnetic fluctuations.
Because some theories suggest that an SC multi phase appears only when in-plane interactions dominate over $c$-axis interactions\cite{S.Khim_Science_2021,T.Yoshida_PRB_2012}, the two-dimensional magnetic correlation is considered to be the key feature in CeRh$_2$As$_2$.
Indeed, some band calculations also indicate the quasi-two-dimensional electronic properties in this system\cite{D.Hafner_arXiv_2021,D.C.Cavanagh_arXiv_2021,K.Nogaki_PRR_2021}.

The $XY$-type magnetic anisotropy was also confirmed from the $T_1$ measurements.
In general, $1/T_1$ divided by temperature $1/T_1T$ can be described in terms of fluctuating hyperfine fields perpendicular to the applied magnetic field parallel to the $z$ axis as follow:
\begin{align}
\left(\frac{1}{T_{1}T}\right)_{z} & \propto R_x + R_y, 
\label{eq.3}
\end{align}
where $R_i \equiv |H_{\rm{hf},i}(\omega_{\rm{res}})|^{2}$ and $|X(\omega)|^2$ denotes the power spectral density of a time-dependent random variable $X(t)$.
Because both As sites have 4-fold symmetry of the atomic position, $R_a = R_b$; thus, the relations $(1/T_1T)_{4f,c} = 2R_a$ and $(1/T_1T)_{4f,\perp} = R_a + R_c$ can be applied.
Figure~\ref{Fig.4} (c) shows the temperature dependence of $R_a$ and $R_c$ at the As(2) site in CeRh$_2$As$_2$ as estimated using the above relations.
$R_c$ was small and showed a weak temperature dependence with a broad maximum at $\sim$ 3~K, and $R_a$ continued to increase on cooling, indicating $XY$-type magnetic anisotropy.
This behavior is similar to static magnetic susceptibility.
From the temperature dependence and the anisotropy of $1/T_1T$, it is concluded that CeRh$_2$As$_2$ shows two-dimensional fluctuations with the $XY$-type magnetic anisotropy, thereby satisfying the condition of a unique SC multi phase.

The $XY$-type magnetic anisotropy is often observed in heavy-fermion superconductors.
Sakai $et~al$. investigated the anisotropy of spin fluctuations in many heavy-fermion systems and found that $T_{\rm SC}$ increases with increasing $\Gamma_c/\Gamma_a$\cite{H.Sakai_JPSJsup_2012}.
Here, $\Gamma_i \equiv (k_{\rm B}\gamma_n^2A_{\rm{hf},i}^2/2\pi R_i)^{1/2}$, which is the local magnetic fluctuation energy in the $i$ direction.
It is a quantity that becomes smaller near the quantum critical point.
$k_{\rm B}$ and $\gamma_n$ are the Boltzmann constant and nuclear gyromagnetic ratio, respectively.
In CeRh$_2$As$_2$, $\Gamma_c/\Gamma_a$ was calculated to be $\sim 6$ at 1.5~K, indicating a large $XY$-type anisotropy of spin fluctuations.
According to the relation between $\Gamma_c/\Gamma_a$ and $T_{\rm SC}$ in ref.34, this value corresponds to $T_{\rm SC}$ = 10~K; this is much higher than the actual $T_{\rm SC}$ of $\sim 0.3$~K.
The small $T_{\rm SC}$ in CeRh$_2$As$_2$ might be related to the uniqueness of its crystal structure.

Finally, we comment on the relationship between the magnetic anisotropy and a magnetic structure in the AFM state inside the SC phase.
From our NQR measurements, the $A$-type AFM (in-plane ferromagnetic and inter-plane antiferromagnetic) order with magnetic moments parallel to the $c$ axis or a helical order with in-plane moments are promising candidates for the magnetic structure\cite{M.Kibune_PRL_2021}.
Considering the $XY$-type magnetic anisotropy observed in the present study, a magnetic structure with an in-plane ordered moment would be preferable, but it is not so simple in Ce-based compounds.
Magnetic anisotropy is mainly affected by single-site crystal field effects, but the direction of ordered moments is determined by the interaction between the moments.
As a result, the direction of ordered moments and magnetic easy axis can be different in some Ce-based compounds\cite{S.Araki_PRB_2003,C.Krellner_JCG_2008}.
In addition, in CeRh$_2$As$_2$, another phase transition, which is considered to be the quadrupole-density-wave transition\cite{D.Hafner_arXiv_2021}, exists above the AFM transition temperature, and the order parameter of this state might be coupled with the AFM moments.
Therefore, detailed measurements of low-temperature properties is important for the determination of the AFM structure.

In conclusion, we performed $^{75}$As-NMR measurements to investigate the normal-state magnetic properties of CeRh$_2$As$_2$.
The magnitude and temperature dependence of the Knight shift at the As(2) site indicated  $XY$-type magnetic anisotropy in CeRh$_2$As$_2$.
For spin fluctuations, spatially two-dimensional $XY$-type antiferromagnetic fluctuations were confirmed from the temperature dependence of $1/T_1$.
These results should be crucial for clarifying the origin of the SC multi phase and the special coexistence of superconductivity and magnetism in CeRh$_2$As$_2$.

\section*{acknowledgments}
The authors would like to thank A. Ikeda for experimental supports, and K. Nogaki, Y. Yanase, S. Ogata, Y. Maeno and S. Yonezawa for valuable discussions.
This work was partially supported by the Kyoto University LTM Center, and Grants-in-Aid for Scientific Research (KAKENHI) (Grants No. JP15H05745, No. JP17K14339, No. JP19K14657, No. JP19H04696, No. JP20KK0061, and No. JP20H00130).
C.G. and E.H. acknowledge support from the DFG through grant GE 602/4-1 Fermi-NEST.

\end{document}